\documentclass[letter]{aa} 
\usepackage{graphicx}
\usepackage{txfonts}
\def\R'HK{R'$_{HK}$\ }
\def\muas{$\mu$as\ }
\def\deltaX{$\Delta$x\ }

\begin{document}
   \title{Using the Sun to estimate Earth-like planets detection capabilities.
 \thanks{}  }

   \subtitle{III. Impact of spots and plages on astrometric detection }

   \author{
     A.-M. Lagrange \inst{1} 
      \and
   N. Meunier \inst{1}
  \and
M. Desort \inst{1}
\and 
 F. Malbet \inst{1}
   }

   \offprints{
     A.M. Lagrange,\\
     \email{Lagrange@obs.ujf-grenoble.fr}
   }

   \institute{
UJF-Grenoble 1 / CNRS-INSU, Institut de Planétologie et d’Astrophysique de Grenoble (IPAG) UMR 5274, Grenoble, F-38041 France
   }

   \date{Received December, 20, 2010 / Accepted January, 11, 2011}

   
   \abstract
   {}
   {Stellar activity is a potential important limitation to the detection of low mass extrasolar planets with indirect methods (RV, photometry, astrometry). In previous papers, using the Sun as a proxy, we investigated the impact of stellar activity (spots, plages, convection) on the detectability of an Earth-mass planet in the habitable zone (HZ) of solar-type stars with RV techniques. We extend here the detectability study to the case of astrometry.  }
   {We used the sunspot and plages properties recorded over one solar cycle to infer the astrometric variations that a Sun-like star seen edge-on, 10 pc away, would exhibit, if covered by such spots/bright structures. We compare the signal to the one expected from the astrometric wobble (0.3 $\mu$as) of such a star surrounded by a one Earth-mass planet in the HZ. We also briefly investigate higher levels of activity.
   }
   {
The activity-induced astrometric signal along the equatorial plane has an amplitude of typ. less than 0.2 \muas (rms=0.07 \muas), smaller 
than the one expected from an Earth-mass planet at 1 AU. Hence, for this level of activity, 
 the detectability is governed by the instrumental precision rather than the activity. We show that 
for instance a one Earth-mass planet at 1 AU
 would be detected with a monthly visit during less than 5 years and an instrumental 
 precision of 0.8 \muas. A level of activity 5 times higher would still allow such a detection with a precision of 0.35 \muas.
We conclude that astrometry is an 
attractive approach to search for such planets around solar type stars with most levels of stellar 
 activity. 
}
   {} 

   \maketitle
%

\section{Introduction: }
Stellar activity is now recognized as a potentially strong limitation for the indirect detection of planets. Indeed, spots and bright structures (plages, network) produce brightness inhomogeneities at the stellar surface that affect the photometric, astrometric and radial velocity (RV) signals. The RV signal is also affected by the inhibition of convection in the active area. The amplitude of the activity-related stellar noise depends on the activity pattern and intensity, which is related to the stellar properties (age, temperature, etc). In the cases of young, active stars, or late-type stars, the signal may mimic that of a giant planet with periods of the order of the star rotational period. In the case of less active, solar-type Main-Sequence stars, the noise is much lower, but can nevertheless affect the detection of terrestrial planets. In two recent papers, we have investigated the impact of spots (\cite{lagrange2010}; herefater paper I), plages and convection (\cite{meunier2010}; herefater paper II) on the detectability of an Earth mass planet located in the habitable zone (HZ) of the Sun, as seen edge-on and observed in RV  or in photometry. To do so, we took into account all spots and plages recorded over one solar cycle. We showed that providing a very tight and long temporal sampling (typically twice a week over more than 2 orbital periods) and an RV precision in the 10 cm/s range, the photometric contribution of plages and spots should not prevent the RV detection of Earth-mass planets in the HZ. This is no longer true when convection is taken into account. 
 
Here, we complete the previous work by estimating the astrometric signal produced by the same structures over the same solar cycle, and we compare it to the astrometric signal induced by an Earth-mass planet located at 1 AU from the Sun. We also compare this signal to the RV and photometric signals.
    \section{Data and results}
\subsection{Data}
The data used to compute the astrometric signals are fully described in paper II. In brief, they consist of all (20873) sunspot groups larger than 10~ppm of the solar hemisphere, provided by USAF/NOAA (http://www.ngdc.noaa.gov/stp/SOLAR/) and all (1803344) bright structures : plages (in active regions) and network structures (bright magnetic structures outside active regions) found from MDI/SOHO (\cite{scherrer95}) magnetograms larger than 3~ppm, during the period between May 5, 1996 (JD 2450209) and October 7, 2007 (JD 2454380). The temperatures associated to these two types of structures was deduced from the comparison between the variations of the resulting irradiance and the observed ones (\cite{froehlich98}) during this period. Our temporal sampling is one day, with a few gaps, corresponding to a coverage of 86 $\%$.
  \subsection{Astrometric temporal variations}
\begin{figure}[t]
  \centering
  \includegraphics[angle=0,width=.8\hsize]{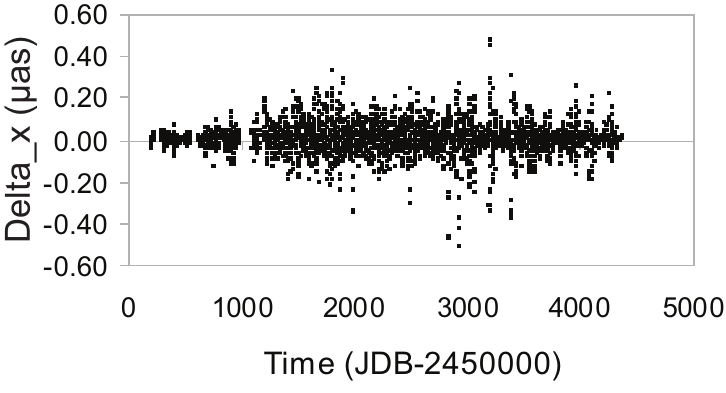} \\
  \includegraphics[angle=0,width=.8\hsize]{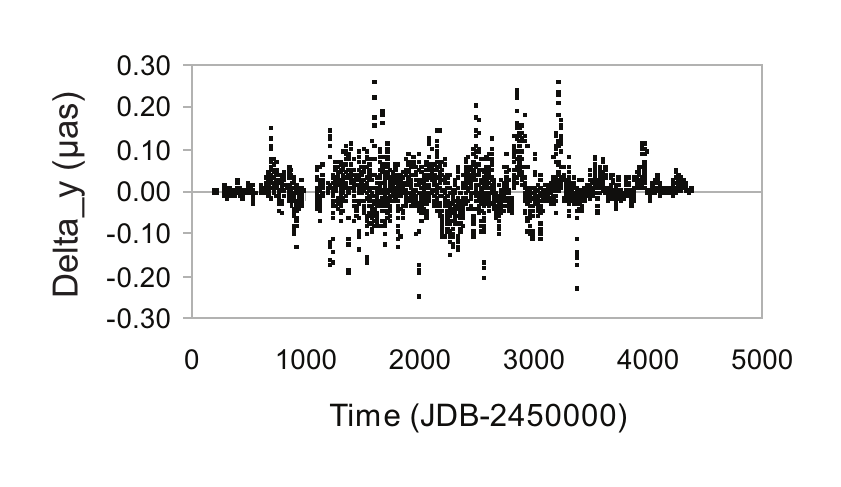}\\
  \caption{Temporal variations of the astrometric shifts along (top) and perpendicular to (bottom) the equator, due to the combination of spots and bright structures. The Sun is supposed to be seen edge-on and located 10 pc away.}
  \label{shift_variations}
\end{figure}

For each date, we compute the position (x,y) of the Sun photocenter due to these dark and bright structures, assuming it is seen edge-on, 10 pc away. The resulting shifts along and perpendicular to the equatorial plane are shown as a function of time in 
Fig.~\ref{shift_variations}. We report in Table~\ref{stats}
the rms of the shifts during the entire cycle, as well as during low activity and high activity period, and the corresponding values for the RV and photometric signals. 

\begin{figure}[t]
  \centering
  \includegraphics[angle=0,width=0.8\hsize]{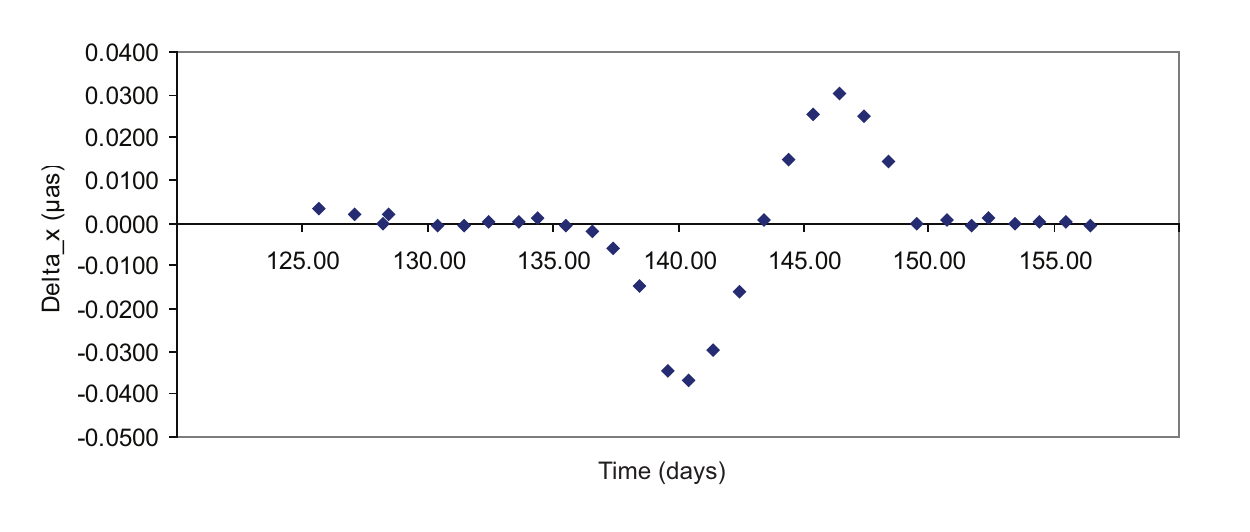}\\
  \includegraphics[angle=0,width=0.8\hsize]{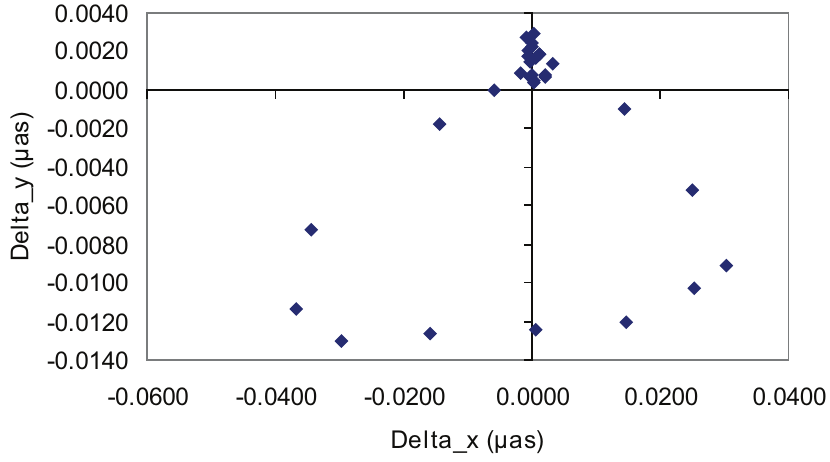}\\
   \caption{Astrometric shift (deltax) as a function of time (top) and excursions of the photocenter (bottom) during a low activity period (JD 2450335 to  JD 2450365).}
  \label{boucle}
\end{figure}

Finally, we provide in Fig.~\ref{boucle} the temporal variations of the astrometric shift along the equator as well as the global excursion of the photocenter during a short 30-day period during the low activity phasis. 
The astrometric signal, if measured precisely enough, allows the determination of the star rotation period. So would also the RV and photomteric signals. The astrometric signal would also provide additional and unique information. In Fig.~\ref{boucle}, we see that the structure at the origin of the observed variations is located on the southern hemisphere of the star, it moving from East to West, ie from the left to right in the lower panel of Fig.~\ref{boucle}. Therefore, the astrometric signal can allow, within one rotation period, to determine the direction of the star rotation axis in the plane of the sky. Note that this assumes that we know by photometry or RV measurements if the perturbating structures are hotter or cooler than the photosphere and that this requires a very good astrometric precision. Also, one needs to isolate and monitor one single structure, so only periods of low activity can provide such information (indeed, during more active periods, the contributions of all individual structures mix up).

\begin{table}[t]
  \caption{Measured rms for the shifts, RV and photometry due to the spots and bright structures}
  \begin{center}
    \begin{tabular}{l l l l l l}
      \hline \hline
      Period  &rms($\Delta$X) &rms($\Delta$Y) & rms(RV) & rms(RV) & rms(TSI)\\ 
      & & & w/o conv. & with conv. & \\ \hline
      
      all	&0.07	&0.05 & 0.33 & 2.4 & 3.6 10$^{-4}$ \\
      high1	&0.09	&0.06 & 0.42 & 1.42 & 4.5 10$^{-4}$ \\
      high2	&0.08	&0.05 & 0.37  &1.62 & 3.9 10$^{-4}$\\
      low	&0.02	&0.01	& 0.08 & 0.44 & 1.2 10$^{-4}$\\
      \hline
    \end{tabular}
\end{center}
\tablefoot{Shifts in x and y are in \muas, RV are in m/s. RV and photometry rms are taken from Paper II.  The entire cycle (refered as "all") as well as low and high activity periods are considered. The low activity period  is  from July 1, 1996 (JD 2450266) to April 1, 1997 (JD 2450540), and the first considered high activity period (refered as "high1") is from February 1, 2000 (JD 2451576) to November 1, 2000 (JD 2451850). Both correspond to the low and high activity 9-months periods considered in Paper II. We also consider a longer (923 days) activity period (refered as "high2"), from JD 2451577 to JD 2452500.      }
\label{stats}
\end{table}

\subsection{Relations between RV, photometric and astrometric signals}
\begin{figure}[t]
  \centering
  \includegraphics[angle=0,width=.8\hsize]{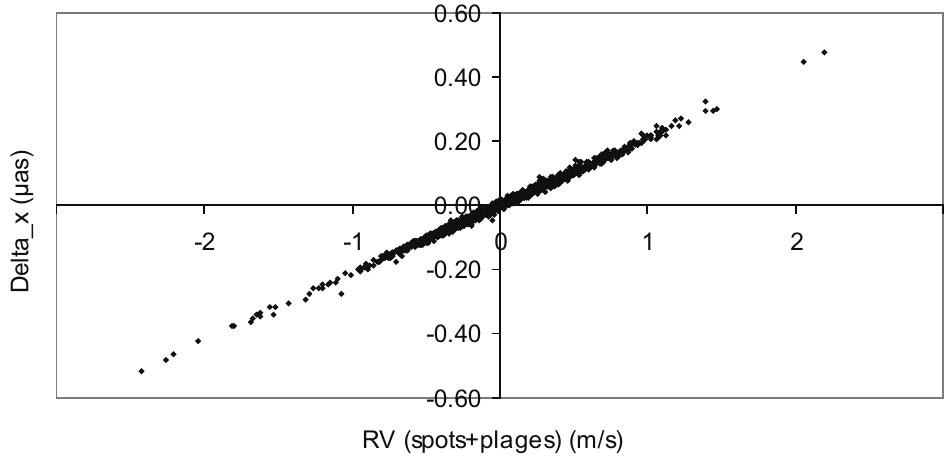} \\
 \includegraphics[angle=0,width=.8\hsize]{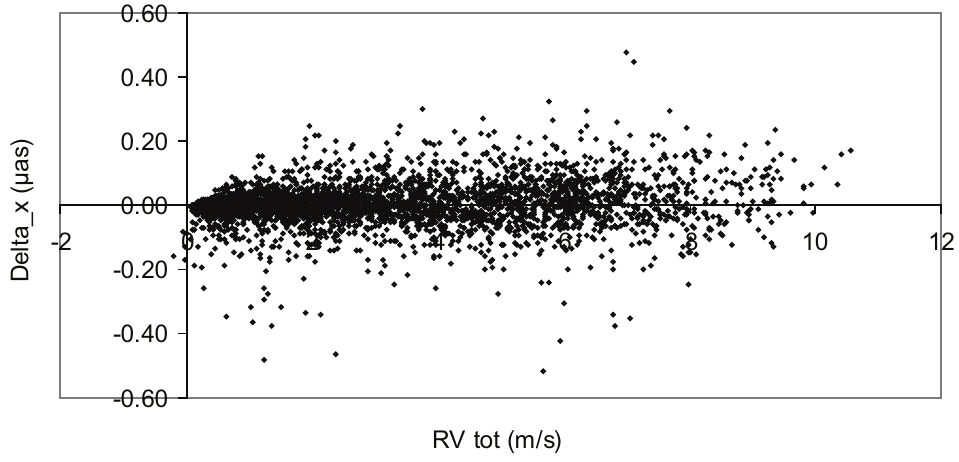} \\
  \caption{Astrometric shifts along the equatorial plane versus RVs due to spots and plages (top) and to spots, plages and convection (bottom). No instrumental noise is assumed here.}
  \label{correl_deltax_rv}
\end{figure}

The astrometric variations along the equatorial plane are well correlated with the RV signal due to spots and plages (see top panel of Fig.~\ref{correl_deltax_rv}). For a single spot, the ratio between the shift and the RV depends on the star's properties ($v.\sin i$) and is almost independant from the spot properties. The correlation found here shows that this property remains when taking several spots/structures into account. The slope of the RV variations (in m/s) to the shifts (in \muas) is  5, compatible with the value expected from a single spot. Once convection is taken into account in the RV computation, however, there is no correlation any longer between the astrometric shift in x and the RV variations (see Fig.~\ref{correl_deltax_rv}). This difference of behaviour between the "no- convection" and "convection" cases could in principle be used to estimate the level of convection, if any, associated to the star.

Finally, no correlation is found between the relative photometric variations and the astrometric shifts (either the \deltaX or the shift to center). This is not surprising as the photometric variations depend marginally on the spots/plages locations, conversely to the astrometric shifts.

\section{Discussion}
\subsection{Planet detection in astrometry}
\begin{figure}[t]
  \centering
  \begin{tabular}{cc}
  \includegraphics[angle=0,width=.45\hsize]{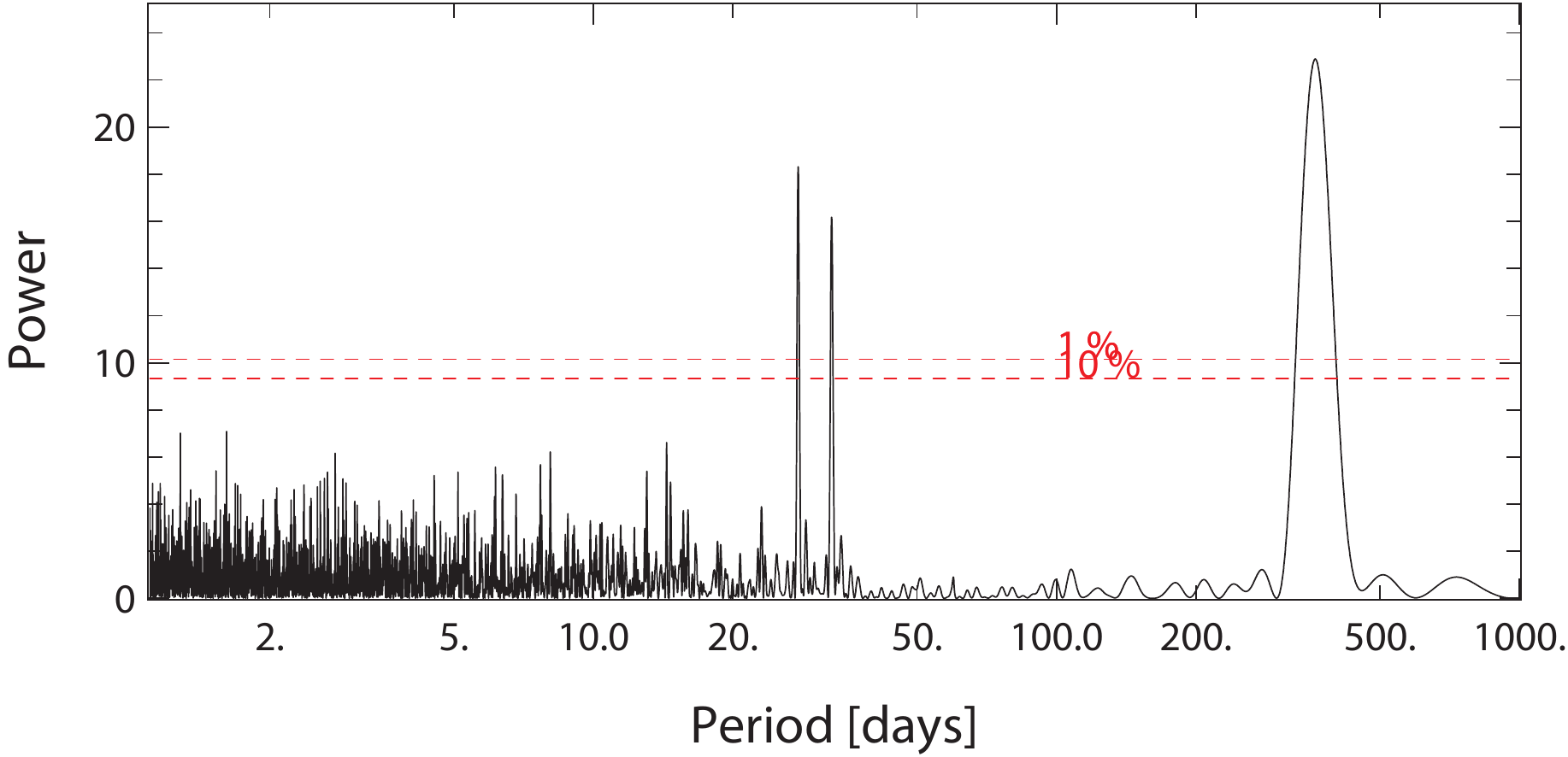}&
  \includegraphics[angle=0,width=.45\hsize]{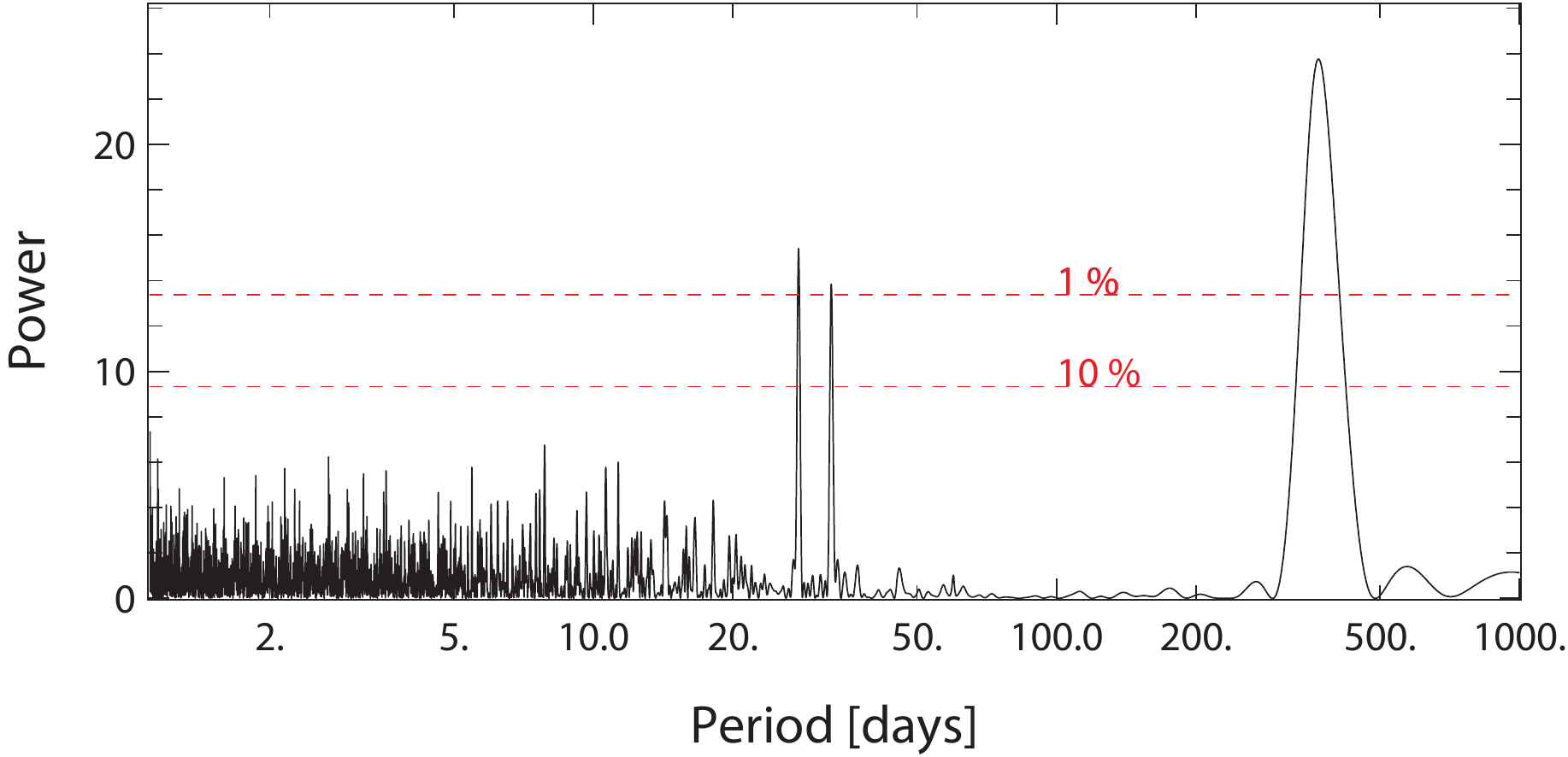}\\
  \includegraphics[angle=0,width=.45\hsize]{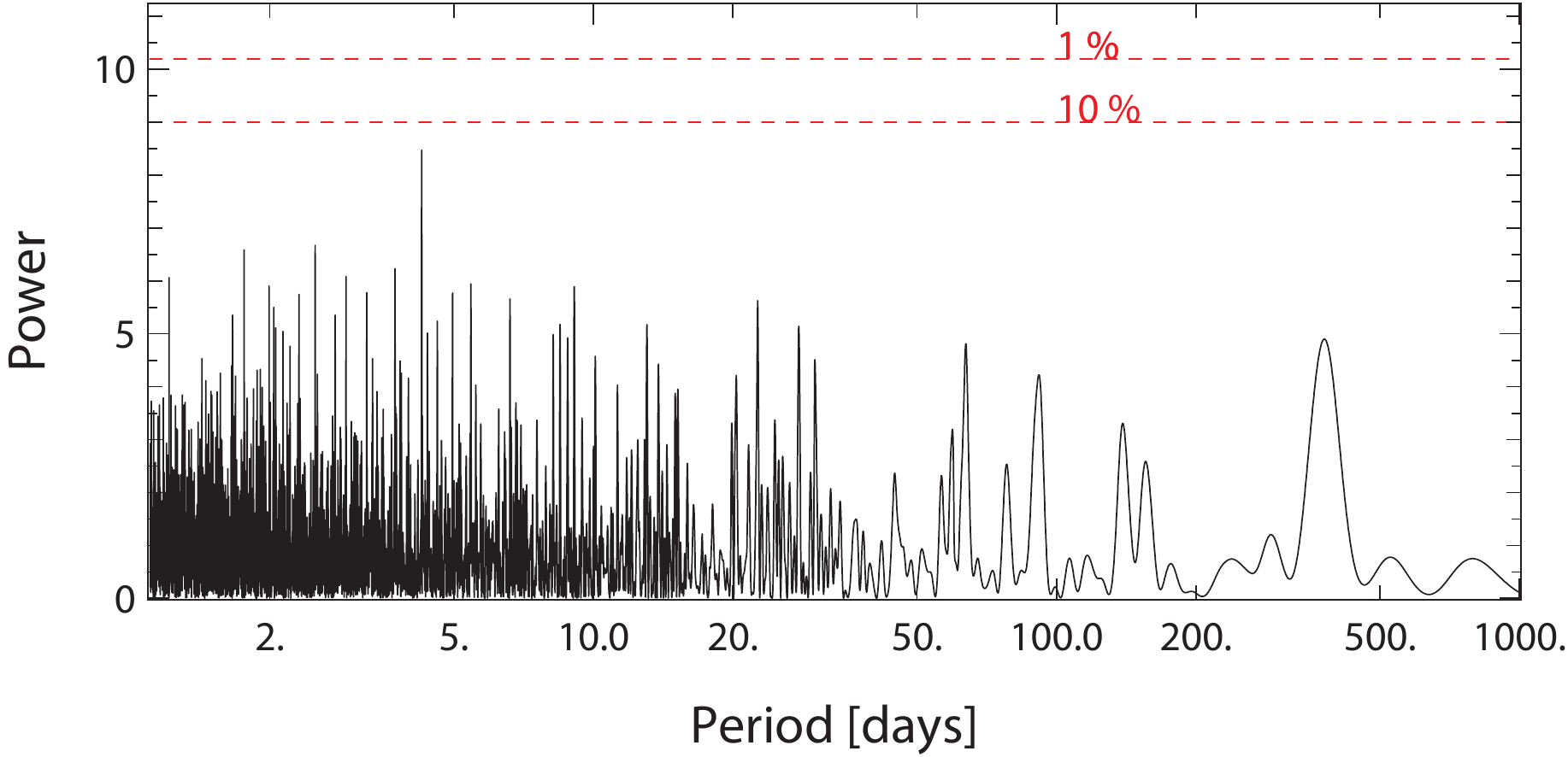}&
  \includegraphics[angle=0,width=.45\hsize]{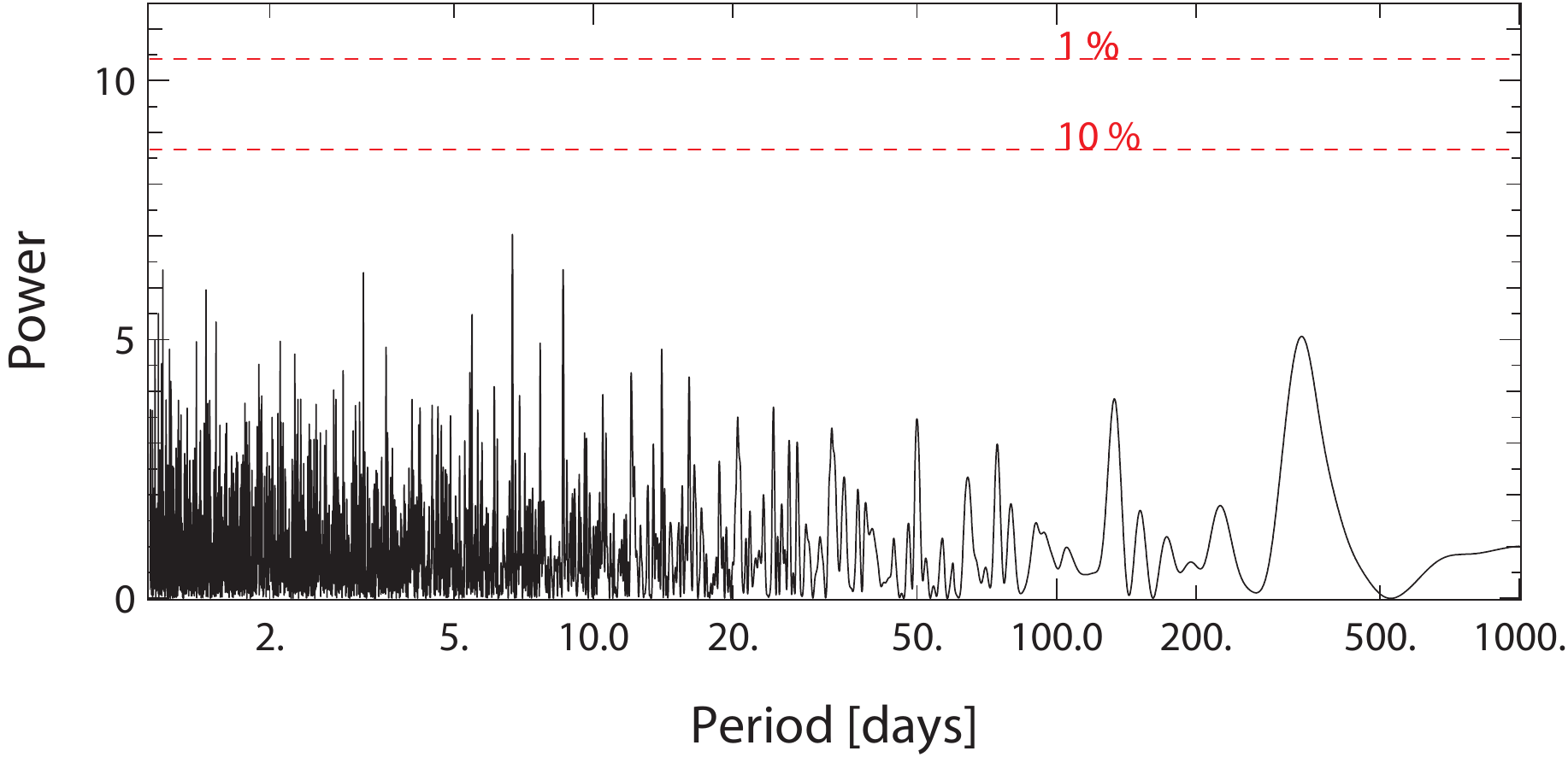}\\
\end{tabular}
  \caption{Periodogrammes of the shift along the equatorial plane of the Sun viewed edge-on and surrounded by a one Earth-mass planets at 1AU, and observed once a month approximately, during 50 months. From top to bottom: 1) low activity period, no instrumental noise; 2) high activity period, no instrumental noise; 3) low activity period, 0.8 \muas rms instrumental noise; 4) high activity period, 0.8 \muas rms instrumental noise.}
  \label{perio}
\end{figure}

We see in Fig.~\ref{shift_variations} that the activity-induced shifts are most of the time smaller than 0.2 $\mu$as, ie smaller than the amplitude of the shifts induced by an Earth mass planet at 1 AU (0.33 \muas along the equatioral plane and 0 \muas perpendicular to the equator). These values are in agreement with the estimations of \cite{makarov2010}. The rms of the equatorial shift over the entire cycle is $\simeq$ 5 times smaller than the amplitude of the Earth signal, and 4 times smaller during high activity. This means that the Sun activity would not prevent to detect an Earth located in the HZ, provided the precision of the data allows such mesurements. This is illustrated in Fig.~\ref{perio} where we show the periodogramme of the astrometric shifts along the equatorial plane once a one Earth-mass planet has been added (no noise). In this example, we use a limited amount of data instead of the whole set of data, in order to be closer to a real case. The chosen temporal sampling is of 1 month $^{+}_{-}$ 5 days over 50 months, typical of the strategy adopted for the NEAT instrument recently proposed to ESA (\cite{leger2010}), dedicated to a systematic astrometric search for exo-Earths in the HZ of nearby stars. 

We then explored different instrumental noises. An example is given in the lower panel of Fig.~\ref{perio} where we assume a noise of 0.8 \muas \footnote{this value corresponds to the precision that NEAT will provide in 1 hour observation. In practice, each visit will be longer (up to $\simeq$ 5 hours) to ensure a better precision (precision is proportional to $\sqrt{t_{\rm obs}}$). }. The temporal sampling is the same as before. The planet is still detectable. However, in that case, the correlation seen in the previous section between the astrometric shift along the equatorial plane and the RV variations is not any longer present. This is due to the fact that the instrumental noise is larger than the activity-induced noise. The consequence is that if we take the instrumental noise into account, for this level of activity, it will not be possible to estimate the level of convection using the astrometric and RV data.

\begin{figure}[t]
  \centering
\includegraphics[angle=0,width=.45\hsize]{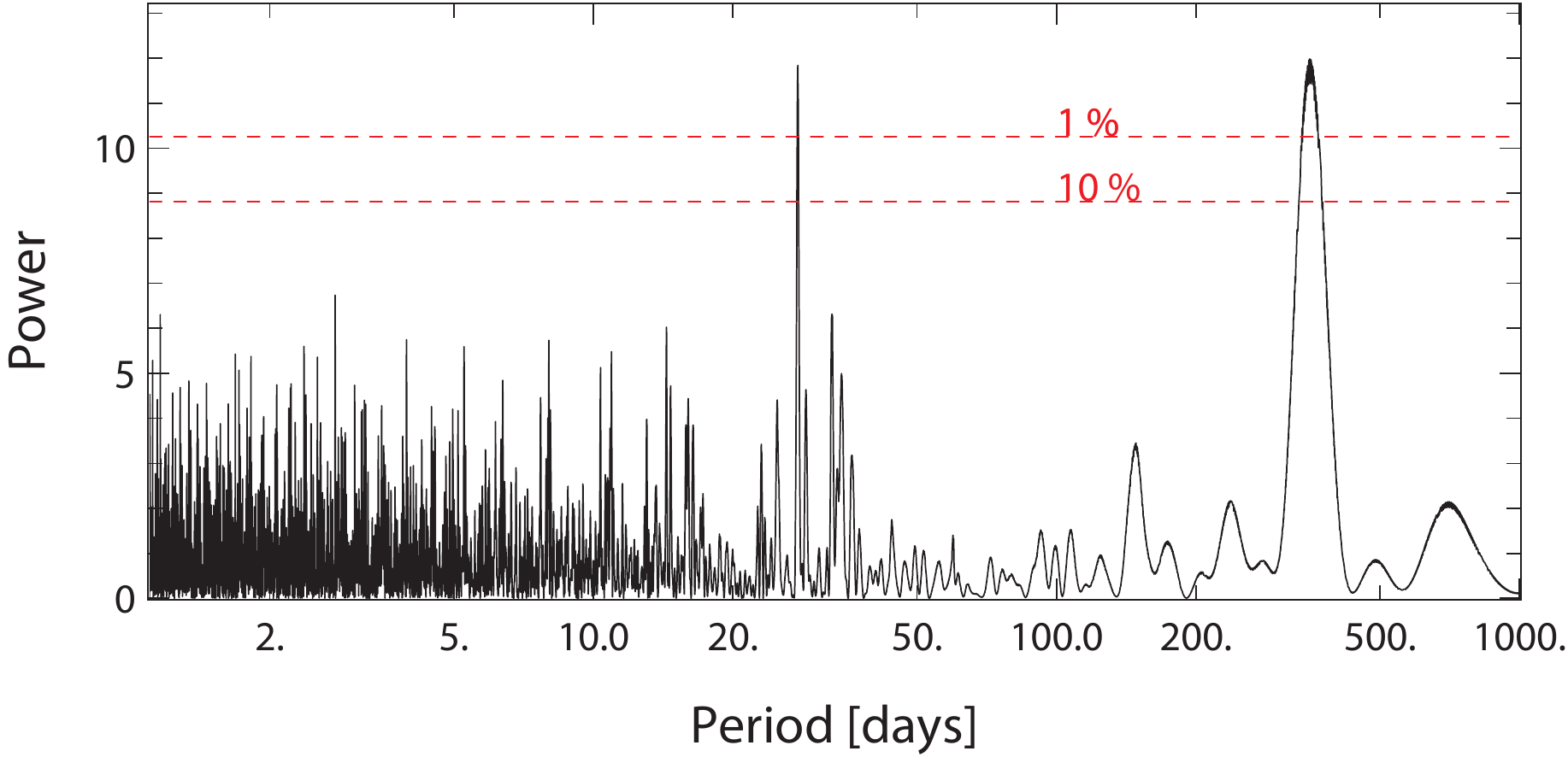} 
\includegraphics[angle=0,width=.45\hsize]{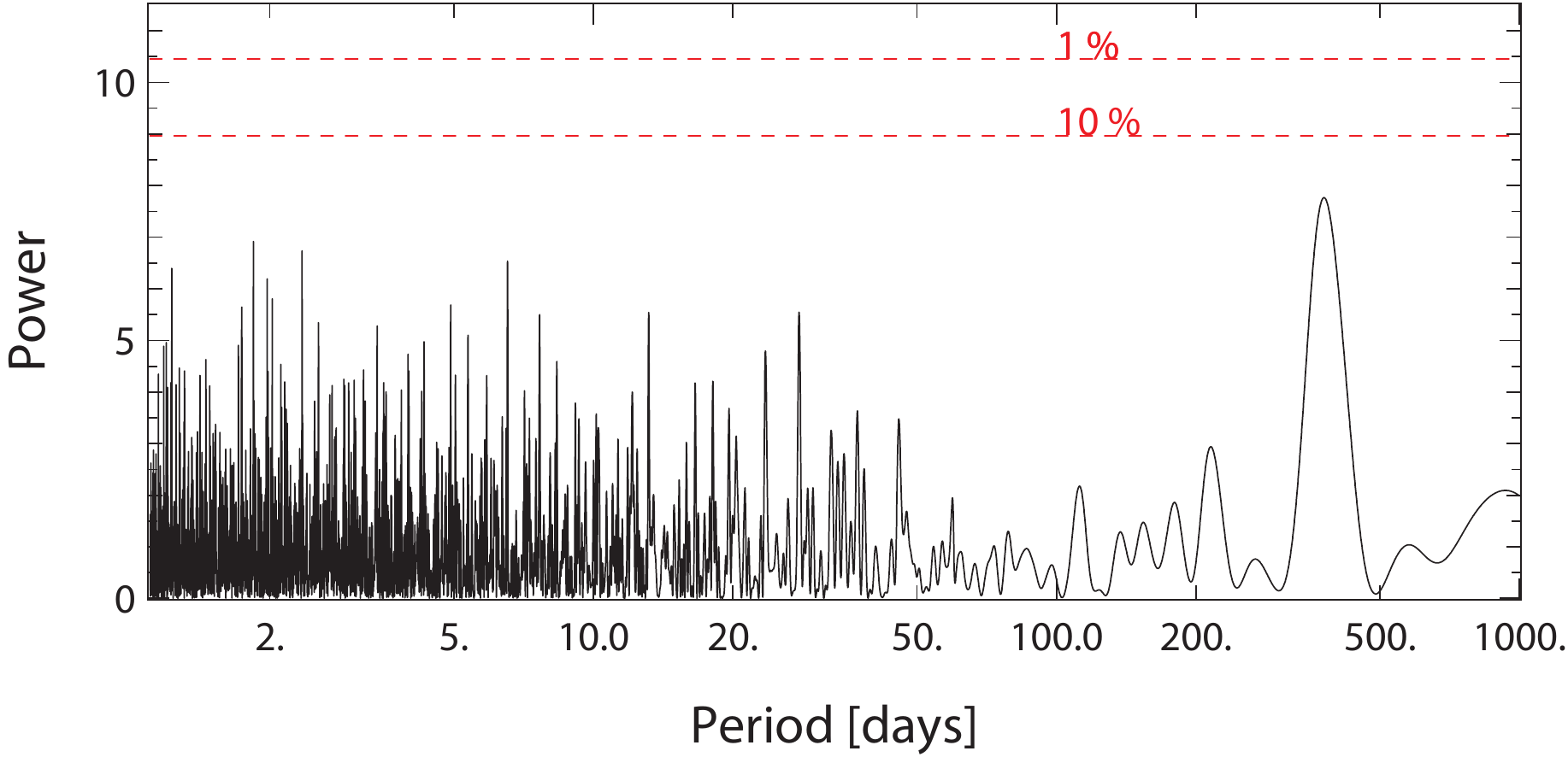}\\
  \caption{Periodogrammes of the shift along the equatorial plane assuming a Sun 5 times more active than observed, viewed edge-on and surrounded by a one Earth-mass planets at 1 AU, and observed once a month approximately, during 50 months. A 0.35 \muas rms instrumental noise has been added to the data. Left: low activity period; right: high activity period.}
  \label{perio_5act}
\end{figure}

\begin{figure}[t]
  \centering
  \includegraphics[angle=0,width=.8\hsize]{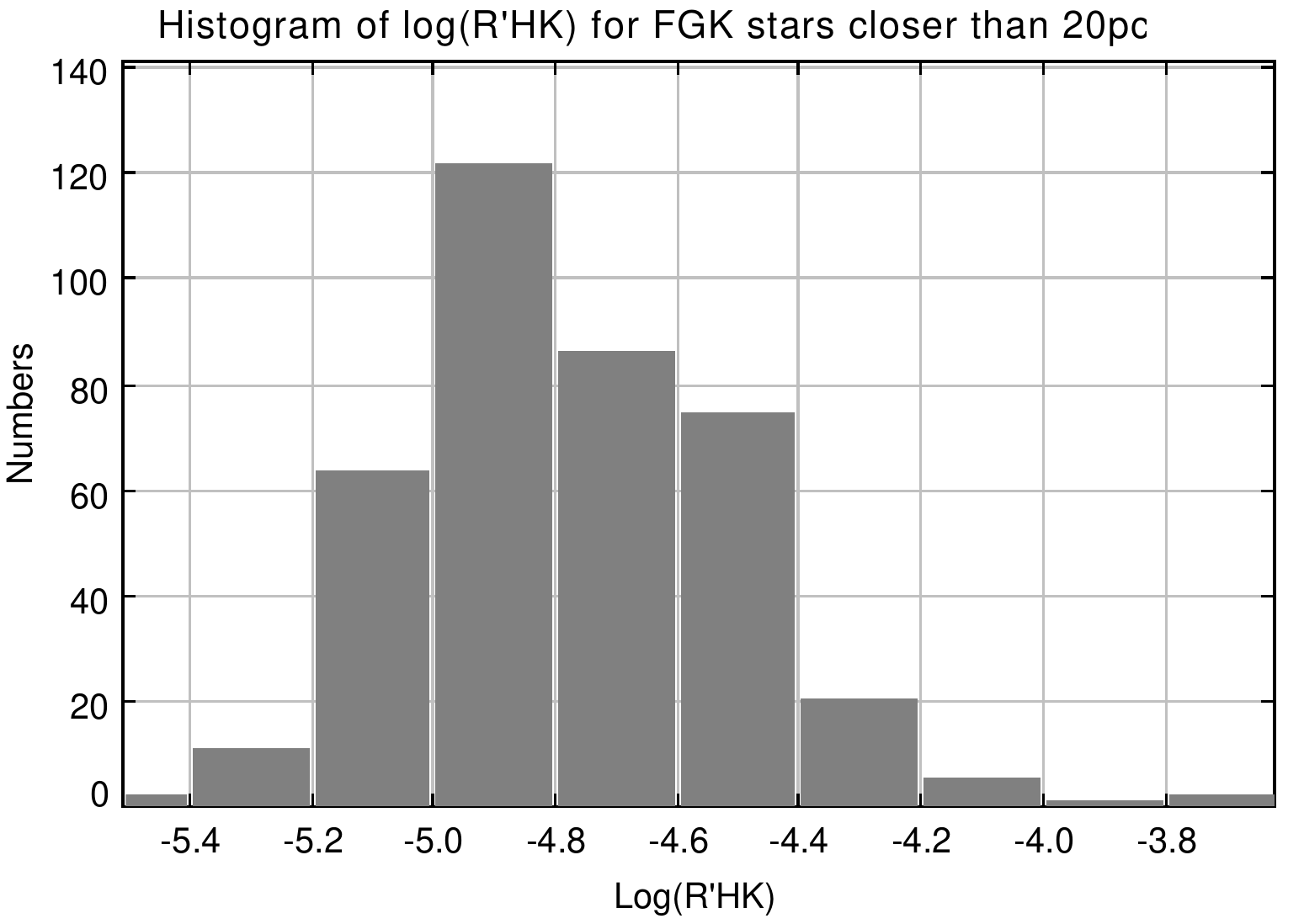}\\
  \caption{Histogram of log(R'HK) for all Hipparcos stars closer than 20 pc, with a known R'HK.}
  \label{stats_hip}
\end{figure}

We finally explore different levels of activity. We found that an instrumental precision similar to that expected for the NEAT instrument for a 5 hours-long visit of a G2 star located 10 pc away (0.35 \muas) will easily allow the detection even for stars 5 times more active than the Sun (we assume conservatively that such a star will produce shifts five times higher than the Sun) \footnote {in practice, a level of activity five times higher will probably induce shifts smaller than 5 times the shifts, as statistically, there will be more chances that the effects of structures located on a given side of the hemisphere be balanced by that those of structures on the opposite side}.
This is illustrated in Fig.~\ref{perio_5act} where we show the periodogrammes of the shift along the equatorial plane (same sampling as before) during the low and high activity periods. Given \cite{lockwood07} relation between the relative photometric variations and the Ca H \& K line strength index log(R'HK), and given the log(R'HK) for the Sun (between $\simeq$ $-$4.85 and $-$5.0 respectively during the high and low activity phases), this corresponds to a log(R'HK) smaller than $-$4.25. 
Most of active stars fall into this regime. Indeed, among the 385 Hipparcos stars closer than 20 pc with a known R'HK, only 8 have log(R'HK) $\geq-$4.2 (see Fig.~\ref{stats_hip}; \cite{maldonado10})).

\begin{figure}[t]
  \centering
\includegraphics[angle=0,width=.8\hsize]{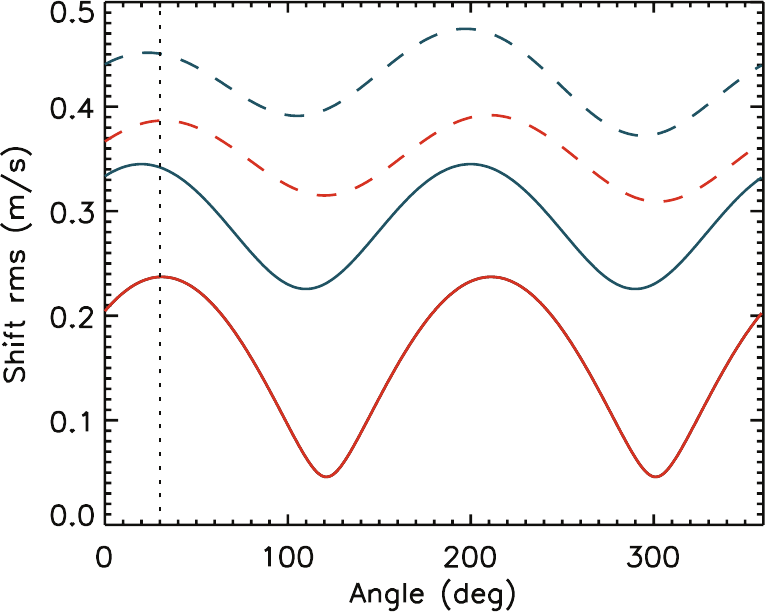}\\
  \caption{Rms of the shift along the x-axis of a rotating referential as a function of its rotational angle (see text). In red,  case of solar activity. In blue, case of 5 times the solar activity. In all cases, we consider that the instrumental noise is 0 (plain lines) or 0.35 microas (dashed lines). The vertical dashed line shows the angle of the inclination of the planet orbital plane. }
  \label{fig_angle}
\end{figure}

So far we have assumed that the angle of the planet orbit (supposedly identical to the equatorial plane) is known. In the case of real observations though, this information will not be known a priori and the planet may orbit outside the star's equatorial plane. It appears that even in the presence of stellar activity and instrumental noise at a level of 0.35 \muas per measurement, the astrometric data will allow to recover this information. We consider the example of a planet on an orbit inclined by -60 degrees with respect to the North. The star is supposed to have either the same activity level as the Sun, or 5 times this level. The astrometric shifts are measured in a referential which is allowed to rotate from 0 (equatorial plane) to 360 degrees. We measure in each case the shifts projected on this referential, and the associated rms along each axis of the referential. We consider either no or 0.35 \muas level instrumental noise, and either a solar-like level of activity or 5 times this level. The temporal sampling is the same as before (50 data points). We show in Fig.~\ref{fig_angle} the rms as a function of the rotation angle of the referential. We see a clear maximum at an angle equal to the planet inclination. Hence the position angle of the planet orbit can be retrieved.
 
We conclude that in most cases, stellar activity will not be the dominant factor for the planet detectability in astrometry. The dominant factor will rather be the instrumental precision. Note that if present, other larger planets will also induce noises, which are not considered in this paper, but the series of blind tests conducted to estimate the detection capabilities  of Earth-like planets in multiple systems by space-borne astrometry (\cite{traub2010}) have showed that this noise can be well circumvented. 

\subsection{Comparison to RV data}
The situation is then different from the RV data, as the rms of the RV due to spots and bright structures only (i.e. convection not taken into account) during resp. the entire cycle, the low activity period and the high activity periods are 3, 1 and 4 times larger than the amplitude of the Earth RV signal (see Table~\ref{stats}).  When taking convection into account, the corresponding ratio become 24, 4 and 14, respectively. Therefore, for RV techniques, the detection limit is set by the stellar activity (and mainly by convection effects) rather than the instrumental precision (provided precisions of 10 cm/s are available). 

To prepare and accompany a space astrometric misson dedicated to Earth-mass planet detection, a number of previous RV, or photometric monitoring would certainly be very precious. In particular, RV measurements are very powerful to derive the properties of the star's activity. They are in fact more powerfull than astrometry as the rms of the activity-induced RV variations is much larger than the instrumental precision (conversely to astrometry). Previous long-term RV monitoring can allow to identify the presence of larger bodies in the system and characterize them (note that lower precision astrometric monitoring can also do this). During the astrometric measurements, simultaneous RV observations with a precision of about 10 cm/s would provide detailed information on the actual activity of the star and will help the analysis of the astrometric data. 

\begin{acknowledgements}

  We acknowledge support from the French CNRS. We are grateful to 
  Programme National de Plan\'etologie ({\small PNP, INSU}). We also thank Alain L\'eger and Mike Shao for fruitfull discussions on the  NEAT project and the referee for his prompt comments.
 
\end{acknowledgements}


\begin{thebibliography}{}
%
\bibitem[Froehlich and Lean 1998]{froehlich98} Froehlich, C, \& Lean, J., 1998, Geophys. Res. Letters, 25, 4377
%
\bibitem[Lagrange et al. 2010]{lagrange2010} Lagrange, A.-M., Desort, M., \& Meunier, N., 2010, A$\&$A, 512, A38
%
\bibitem[L\'eger 2010]{leger2010} L\'eger, A., 2010, private communication
%
\bibitem[Lockwood et al. 2007]{lockwood07} Lockwood, G. W., Skiff, B. A., Henry, G. W. et al., 2007, ApJ, 171, L260
%
\bibitem[Makarov et al. 2010]{makarov2010} Makarov, V. V., Parker, D., Ulrich, R. K., 2010, ApJ, 717, 1202
%
\bibitem[Maldonado et al. 2010]{maldonado10}  Maldonado, J., Mart{\'{\i}}nez-Arn{\'a}iz, R.~M., Eiroa, C., Montes, D., Montesinos, B., 2010, A$\&$A, 521, A12
%
\bibitem[Meunier et al. 2010]{meunier2010} Meunier, N., Desort, M. \& Lagrange, A.-M., 2010, A$\&$A, 512, A39
%
\bibitem[Scherrer et al. 1995]{scherrer95} Scherrer, P. H., Bogart, R. S., Bush, R. I., et al., 1995, Sol. Phys., 162,129
%
\bibitem[Traub et al. 2010]{traub2010} Traub, W., Beichman, C., Boden, A.F., et al, 2010, EAS Publications Series, 42,191
\end{thebibliography}
\end{document}